# Experimental Evidence of a Bonded Dineutron Existence

I.M. Kadenko[*]

*International Nuclear Safety Center of Ukraine; Department of Nuclear Physics*
*Taras Shevchenko National University of Kyiv, 01601, Kyiv, Ukraine*

Experimental observation of $^{159}$Tb$(n,^2n)$ reaction product was performed with application of the activation technique. Tb specimen of natural composition was irradiated with $(d,d)$ neutrons of 5.39 and 7 MeV energy at the AMANDE neutron generating facility. Several instrumental spectra of Tb specimen were measured with HPGe spectrometer in 1.5 years after last irradiation. An unexpected 944.2 keV γ-ray peak was observed. Other γ-ray lines of $^{158}$Tb were identified as well. A bonded dineutron emission with the binding energy ($B_{dn}$) within limitations 1.3 MeV $< B_{dn} <$ 2.8 MeV is evidenced by the energy of incident neutrons and by $^{158}$Tb presence in output channel. The specific nuclear properties of $^{158}$Tb as deformed nucleus were then discussed to explain a bonded dineutron formation with certain half-lives based on theoretical assumptions and corresponding calculations, using standard parameters for this mass region.

**Introduction.** The purpose of this Article is to point out that the dineutron may exist as a bonded particle in the vicinity of the heavy nucleus in output channel of nuclear reaction. The field of nuclear physics knows long history of searching for dineutron bound states. Numerous attempts had been made to look for the dineutron either as a structural component of light nuclei (the two-neutron halo) or as a product of nuclear reactions on heavier nuclei, see [1 - 3], and references therein. These works covered atomic mass regions $A \leq 65$ and $A \geq 209$ for a dineutron search and detection but there were no attempts to link a research of rare earths and the dineutron physics. At the same time many papers claim the dineutron itself or some virtual dineutron states may exist in the presence of neutron-rich nuclei, but none of them prove whether such states correspond to namely dineutron bound states. I assumed that in case the dineutron should exist, it would be expected to be produced in $(n,^2n)$ reaction on heavy and neutron-rich nuclei over a range of neutron incident energy just below the threshold energy for the corresponding $(n,2n)$ reaction. This idea is not inconsistent with the work due to Migdal [4]. He suggested that while the dineutron is known to be unbound, in the field of the heavy nucleus it may exist as a bonded particle near the nuclear surface of this heavy nucleus. In this study our attention was focused on Terbium (Tb, $A=158 \div 160$) as an element of the rare earth group based on our previous research [5, 6]. A reason to select Tb is its special nuclear properties, such as strong deformation in ground state. Suggestions for further nuclear physics research of the rare earths may include such exciting subjects as the search for the nuclear limits of stability and proton radioactivity [7], as well as new reaction channels.

**Experimental technique.** To study possible new reaction channels with expected low activities in output channel, the neutron activation technique is among the most suited ones. In order to understand the behavior of new neutron induced reactions on Tb, the irradiations of specimen were made with incident neutrons at the AMANDE facility (the Institute for Radiation Protection and Nuclear Safety (IRSN), Cadarache) [8], which is based on a HVEE 2 MV Tandetron accelerator system. Neutrons were generated using the nuclear reaction between accelerated deuterons and a thin deuterated target composed of a titanium layer saturated by deuterium attached to 0.5 mm thick silver backing. Neutron fields with energies of 5.39 and 7 MeV were obtained in the direction of the deuterium beam through the $^2$H$(d,n)^3$He reaction. For neutron irradiations one Tb sample was used in a shape of a cylinder of 30 mm diameter with 5 mm thickness (total mass of 28.9 g), which was placed under 0° angle with respect to the incoming deuteron beam. Distance between Tb sample and target varied within 3.3 ÷ 15.3 cm. The mean values of the calculated neutron energy in the specimen were therefore of 5.35 and 6.85 MeV. Total 6.85 MeV neutron fluence delivered at Tb sample was estimated for the case of the three sequential irradiations culminating in $(2.69 \pm 0.08) \cdot 10^{10}$ cm$^{-2}$ with total irradiation time $6.84 \cdot 10^4$ s [9]. For 5.35 MeV total neutron fluence at the specimen was determined as $(2.10 \pm 0.07) \cdot 10^{10}$ cm$^{-2}$ with irradiation time $3.05 \cdot 10^4$ s. IRSN is in charge of neutron metrology activities as a designated institute of the French National Metrological Institute (LNE). The reference neutron fluences and corresponding energies provided for this research are therefore traceable to international standards. Geometry of irradiation is presented in Fig.1. More details are available in our paper [10].

A set of gamma spectrometry instrumental spectra measurements was taken about one plus year after the first and second irradiations were completed, using the coaxial HPGe detector GC2020. The detector was properly shielded with a lead housing and connected to a NIM-based multichannel analyzer. An unexpected 944.2 keV γ-ray peak appeared in the instrumental spectra with the acquisition time ~ $10^6$ s (Fig. 2) only after irradiation in neutron field with 6.85 MeV. It certainly belongs to $^{158}$Tb decay since neither higher energy nor escape peaks were detected in the energy range up to 4 MeV. At first glance this peak is statistically meaningful with peak to background ratio ~ 1.35. To obtain more reliable peak parameters for spectrum of low statistical quality, let's assume for a given region of interest (ROI) that the location of a small peak is approximately known. From here, one can then apply a transformation that maintains the statistical independence of count numbers in adjacent channels and also diminishes a statistical dispersion of count numbers, without distorting a peak shape. One of such possible transformations is as follows: $S_1(i) = 1/2\{S(i) + S(2p-i)\}$, where $S(i)$ - original γ-ray instrumental spectrum (count number) in channel $i$; $S_1(i)$ - the transformed spectrum; and $p$ - the peak location [11]. Results of such a spectrum transformation are given in Fig. 3 and confirm the validity of this peak. Other peaks with energies 780.2, 79.5 and 181.9 keV were identified as well. All of them are presented in Fig. 4 and Fig. 5. Two more expected peaks with energies 98.9 and 962.1 keV were present in spectra but not considered due to overlapping with background lines and 962.3 keV $^{160}$Tb originated peak. Based on all these peaks validated presence and univocal identification a reasonable explanation of this phenomenon must be provided.

* IMKadenko@univ.kiev.ua

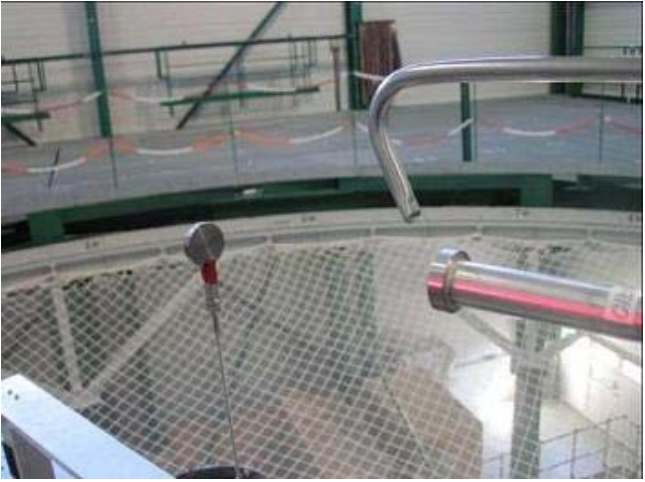

Fig. 1. Irradiation geometry of Tb specimen.

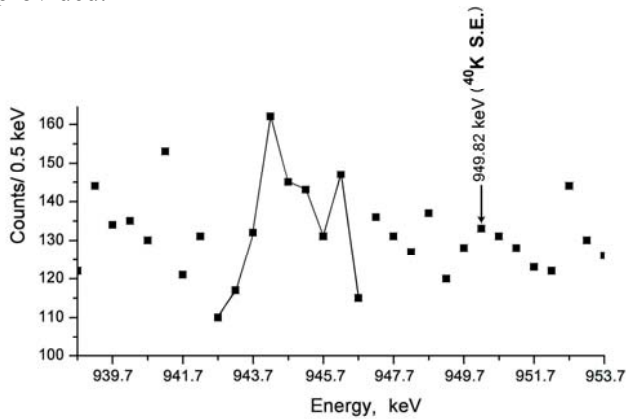

Fig. 2. ROI with 944.2 keV ($I\gamma$=0.439) peak of $^{158}$Gd.

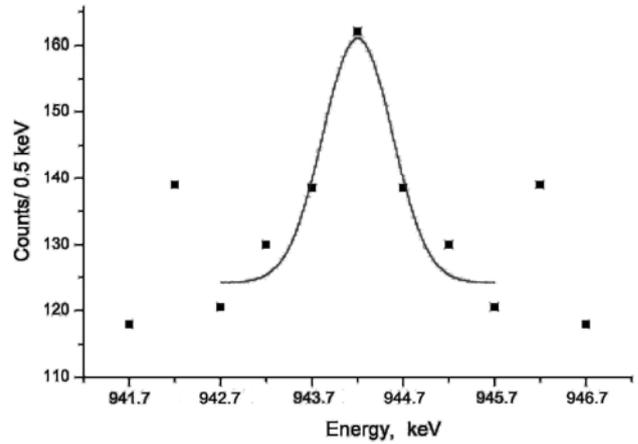

Fig. 3. Transformed spectrum with 944.2 keV peak.

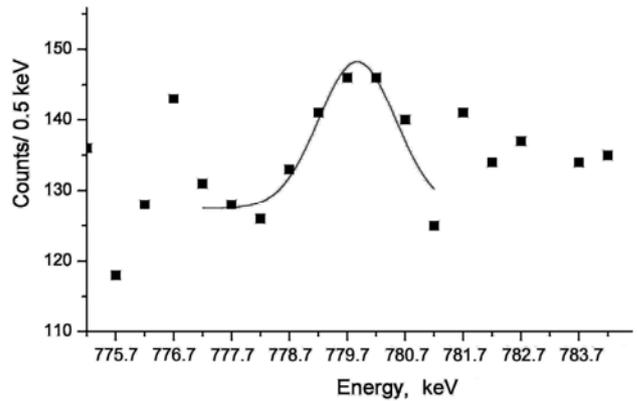

Fig. 4. ROI with 780.2 keV ($I\gamma$=0.0957) peak of $^{158}$Gd.

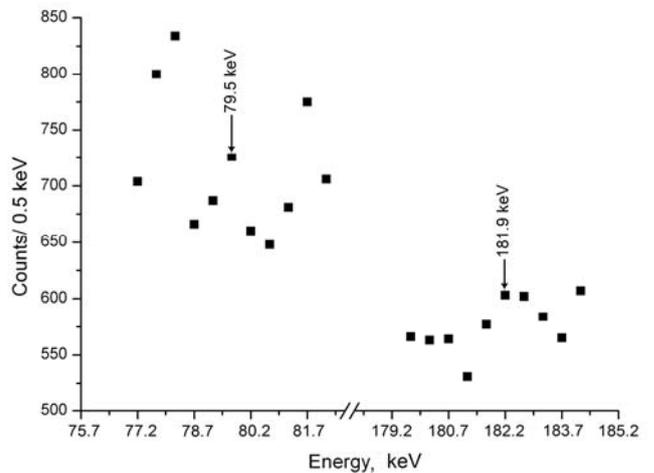

Fig. 5. ROIs with 79.5 keV ($I\gamma$=0.116) and 181.9 keV ($I\gamma$=0.099) peaks of $^{158}$Gd.

**Discussion.** In his paper Migdal assumed and subsequently proved that the strong interaction between the two neutrons, itself insufficient to bond them in a dineutron system may have an additive component caused by the potential well influence of the much heavier nucleus in a close proximity to both neutrons. This leads to the appearance of the additional bound states of the two neutrons, becoming the dineutron in the external field of the massive nucleus. These bound states correspond to the resonance single particle levels at an additional energy branch which terminates at $\varepsilon_c \sim 0.4$ MeV. Then any resonance single particle states are located within the energy range $0 \div 0.4$ MeV. Having these energy bounds set, let's calculate from the expression (25) of Migdal's paper the corresponding boundary atomic masses. Prior to this step, it follows from (22), (22a), and (23) of [4] that both terms under the exponent in (25) must have identical and positive sign and, consequently, a misprint takes place in (25); thus instead of "- $3\pi/2$" we will consider "+ $3\pi/2$". This transforms the expression (25) from [4] into the following one:

$$\varepsilon_c = 2\alpha^2 \cdot EXP[\pi/(2H'_0) + 3\pi/2]. \quad (1)$$

With this amendment it is correct for the spherical shape nuclei to follow the estimates in the latter application of Migdal's paper [4]. Therefore, the energy of resonance state $\varepsilon_c$, at which an additional branch of energy disappears, equals 0.4 MeV and corresponds to atomic mass $A \sim 100$. The upper estimate of $A$ for a bonded dineutron existence gives $A \sim 200$ based on assumption that in free space the dineutron is unbound by about 0.1 MeV [12]. Then the atomic masses of the massive spherical nuclei to comprise a system the heavy nucleus plus the dineutron must be within 100 and 200 u, and Tb fits very well into this atomic mass interval. Furthermore it is important to calculate $\varepsilon_c$ for $^{158}$Tb and, considering the dineutron to double check that a resonance level may have the energy higher 0.1 MeV. From amended expression (1) above for $\alpha = 1/R_z$;

$R_z = (1 + \sqrt{5/(4\pi)} \cdot \beta) \cdot r_0 \cdot A^{1/3}$; $r_0 = 1.25 fm$;

$\beta = 0.271$; $H'_0 = -2.5 A^{-1/3}$; $A = 158$ we get $\varepsilon_c = 0.119$ MeV. Should a resonance state exist within the energy interval $0.100 \div 0.119$ MeV for such system according to Migdal, the dineutron must be bound. As it was stated above, $^{158g+m}$Tb is a deformed nucleus of a prolate shape. Therefore an energetically favorable simultaneous escape of the two paired up neutrons from the Fermi level of $^{160}$Tb most probably occurs at a peripheral part of $^{160}$Tb prolate nucleus in the direction of a polar diameter to further form $^{158g+m}$Tb plus the dineutron near the polar surface of $^{158g+m}$Tb. For the determination of the binding energy of the dineutron from the above described experiment it was only possible to set the upper limit by subtraction from the threshold energy value of the $^{159}$Tb$(n,2n)^{158g+m}$Tb nuclear reaction $E_{th}$=8.18 MeV [13] of the neutron energy 5.35 MeV for which no emission of bonded dineutron was observed. Analogously the lower limit of $B_{dn}$ was estimated by subtracting 6.85 MeV from the $E_{th}$. Finally we get 1.3 MeV<$B_{dn}$<2.8 MeV. These limits are fully consistent with similar ones in [14] and some other expectations in [15]. With these binding energy limits set one can make an estimate for the half-life of the dineutron ($t_{1/2}$) based on the reasonable assumption that the transition $^2n \to d$ occurs via $\beta^-$-decay [14]. Because of low atomic mass of the dineutron ($A_{dn}$=2) this transition characterized by very low value of comparative half-life $f_{dn} \cdot t_{1/2}$ and might occur as superallowed. This gives:

$$f_{dn} \cdot t_{1/2} = 3 \div 3.5 \text{ with}$$
$$lg f_{dn} = 4.0 \cdot lg E_{max} + 0.78 + 0.02 A_d - 0.005(A_d - 1) \cdot lg E_{max}, \quad (2)$$

where $A_d$ - atomic mass of the deuteron; $E_{max}$ – the limiting kinetic energy of $\beta^-$-decay spectrum, in MeV [16]. The results of dineutron half-life calculations from (2) are presented in Table 1.

Table 1. Results of dineutron half-lives calculation.

| $f_{dn} \cdot t_{1/2}$ \ $E_{max}$ | 1.3 | 2.8 |
|---|---|---|
| 3.0 | 0.1592 s | 0.0074 s |
| 3.5 | 0.1857 s | 0.0087 s |

From Table 1 we obtain 0.0074 s <$t_{1/2}$<0.1857 s. By comparing these half-life limitations to $t_{1/2}$ results in [14] one can point out several orders of magnitude difference and expect millisecond rather than second values as the best estimate for the dineutron half-life. With regard to detecting the dineutron it would be most likely impossible to carry out a direct detection using traditional or up-to-date neutron detectors. That is why since the mid of last century as one of the most appropriate methods the one of nuclear reactions was considered with $^{25}$Na or $^{65}$Cu or $^{209}$Bi etc. and the dineutron in the input channel [1, 14]. Such approach may look promising if necessary cross section estimates will be available and a specimen with specific properties could be developed. But prior to this, and based on some features of a dineutron formation from above discussion, this method is unlikely expected to provide any reasonable evidence of a dineutron detection. Then the task to directly detect the dineutron is getting even more complicated and would require to search for correlated two-neutron emission after a dineutron disintegration in a field of bombarding neutrons, or to identify some extra deuterons after dineutrons decay having in mind that $(n,d)$ reaction channel may also be open. Ultimately, the existence of the dineutron as a bonded particle with certain half-life in the very early moments of the Big Bang when nuclear interaction might have been stronger [17, 18] could potentially influence the Big Bang Nucleosynthesis (BBN) [15]. Thus if $B_{dn}$ is even a bit greater than a deuteron binding energy (what is not excluded from our energy limitations above), then nucleons may follow new reaction channels. One of them, $^7$Be$(^2n,n\alpha)^4$He could play a significant role to resolve the lithium problem in BBN. A bonded dineutron, perhaps with some other assumptions, could be helpful also in explaining low abundances of Boron that is observed in old stars but not predicted by BBN yet. Another positive outcome of a bonded dineutron existence could give a chance for more neutrons to survive and could lead to much larger deuteron abundances than in the standard BBN.

**Summary.** In this Article it was shown that the dineutron as a bonded system can exist near the nuclear surface of $^{158}$Tb as a product of ($n,^2n$) reaction on $^{159}$Tb. The assumption above of a dineutron formation is justified by the prediction of Migdal and by the fact that Tb is a strongly deformed nucleus and by the lack of explicit theory of deformed nuclei to fully explain this nuclear reaction mechanism. To make sure such nuclear reaction process is valid, a not so difficult check out may be performed by irradiation during reasonable time of small thickness Tb specimen with 6.5÷7.0 MeV high neutron flux and further detection of 110.3 keV gamma-rays or the prominent K X-rays of 45 keV energy due to 10.7 s isomeric transition in $^{158m}$Tb nuclei, provided a proper detector is selected and the background conditions are satisfactory. Then more precise experiments are needed in the nearest future to study this phenomenon in details. Also one can assume that similar results of a bonded dineutron observation might be expected at least for some other rare earth nuclei.

**Acknowledgment.** The author gratefully acknowledges all of the team members of AMANDE crew for stable and reliable operation of the accelerator.

[1]. Cohen, B.L. & Handley, T.H. An Experimental Search for the Dineutron. Part I: Search for a Stable Dineutron. *Oak Ridge National Laboratory Report No.1832.* (1952); An Experimental Search for a Stable Dineutron. *Phys. Rev.* **92** (1), 101 (1953).

[2]. Strizhak, V.I., Gurtovoy, M.E., Leshchenko, B.E., Prokopets, G.A. & Sitko S.P. *Physics of Fast Neutrons* (Atomizdat, Moscow, 1977) pp. 208-213 [in Russ.].

[3]. Seth, K.K. & Parker, B. Evidence for dineutrons in extremely neutron-rich nuclei. *Phys. Rev. Lett.* **66** (19), 2448 (1991).

[4]. Migdal, A.B. Two interacting particles in a potential well. *Yad. Fiz.* **16**, 427 (1972) [*Sov. J. Nucl. Phys.* **16**, 238 (1973)].

[5]. Dzysiuk, N., Kadenko, I., Köning, A. & Yermolenko, R. Cross sections for fast-neutron interaction with Lu, Tb, and Ta isotopes. *Phys. Rev. C* **81**, 014610 (2010).

[6]. Dzysiuk, N., Kadenko, A., Kadenko, I. & Primenko, G. Measurement and systematic study of (n,x) cross sections for dysprosium (Dy), erbium (Er), and ytterbium (Yb) isotopes at 14.7 MeV neutron energy. *Phys. Rev. C* **86**, 034609 (2012).

[7]. Ivascu M. et al. & Bonetti R. Proton radioactivity in light rare-earth deformed nuclei. *Romanian Reports in Physics*, Vol.**57**, No.**4**, 671 (2005).

[8]. Gressier V. *et al.* AMANDE: a new facility for monoenergetic neutron fields production between 2 keV and 20 MeV. *Radiat. Prot. Dosim.* **110** (1–4), 49 (2004).

[9]. Gressier, V. INSC terbium samples irradiation. *Institute for Radiation Protection and Nuclear Safety Report No. PRP-HOM/SDE/LMDN/2013-327.* (2013).

[10]. Dzysiuk, N., Kadenko, I., Gressier, V. & Köning, A.J. Cross section measurement of the $^{159}$Tb(n,γ)Tb$^{160}$ nuclear reaction. *Nucl. Phys.* A**936**, 6 (2015).

[11]. Zlokazov, V.B. Mathematical methods of the analysis of experimental spectra and spectrum-like distributions. *Sov. J. Part. Nucl.* **16** (5), 1126 (1985) [in Russ.].

[12]. Hammer H.-W. & König, S. Constraints on a possible dineutron state from pionless EFT. *Phys. Lett. B* **736**, 208 (2014).

[13]. http://www.nndc.bnl.gov/qcalc/

[14]. Feather, N. Properties of a Hypothetical di-Neutron. *Nature* **162**, 213 (1948).

[15]. Kneller J.P. & McLaughlin, G.C. Effect of bound dineutrons upon big bang nucleosynthesis. *Phys. Rev. D* **70**, 043512 (2004).

[16]. Friedlander, G., Kennedy J.W., Macias E.S. & Miller J.M. *Nuclear and Radiochemistry* (John Wiley & Sons, Inc., New York, 1981) p.84.

[17]. Coc A., Nunes, N. J., Olive, K. A., Uzan, J.-P. & Vangioni E. Coupled variations of fundamental couplings and primordial nucleosynthesis. *Phys. Rev. D* **76**, 023511 (2007).

[18]. MacDonald J. & Mullan, D.J. Big bang nucleosynthesis: The strong nuclear force meets the weak anthropic principle. *Phys. Rev. D* **80**, 043507 (2009).